\documentclass{PoS}
\usepackage{lineno}

\title{Future prospects of testing Lorentz invariance with UHECRs}

\ShortTitle{Lorentz Invariance Violations in UHECRs}

\author{\speaker{Denise Boncioli}$^{1}$, Armando di Matteo$^{2}$, Francesco Salamida$^{3}$, Roberto Aloisio$^{4,5}$, Pasquale Blasi$^{4,5}$, Piera L. Ghia$^{6}$, Aurelio F. Grillo$^{1}$, Sergio Petrera$^{2,4}$, Tanguy Pierog$^{7}$\\
       $^{1}$INFN/Laboratori Nazionali Gran Sasso, Assergi (L'Aquila), Italy\\
       $^{2}$INFN and Department of Physical and Chemical Sciences, University of L'Aquila, L'Aquila, Italy\\
       $^{3}$INFN and University of Milano Bicocca, Milano, Italy\\
       $^{4}$INFN/Gran Sasso Science Institute, L'Aquila, Italy\\
       $^{5}$INAF/Osservatorio Astrofisico di Arcetri, Firenze, Italy\\       
       $^{6}$LPNHE, Universit{\'e}s Paris 6 et Paris 7, CNRS-IN2P3, Paris, France\\
       $^{7}$Karlsruhe Institute of Technology - Campus North - Institut f$\ddot{u}$r Kernphysik, Karlsruhe, Germany\\
E-mail: \email{denise.boncioli@lngs.infn.it}
}

\abstract{
In the last years a general consensus has emerged on the use of ultra-high energy cosmic rays (UHECR) data as a powerful probe of the validity of special relativity. This applies in particular to the
propagation of cosmic rays from their sources to Earth, responsible for energy suppressions due to
pion photoproduction by UHE protons (the Greisen-Zatsepin-Kuzmin limit) and photodisintegration
of UHE nuclei (the Gerasimova-Rozental limit). A suppression in the flux of UHECRs at energies
above 40 EeV -- as expected from both these interactions -- has been established experimentally beyond
any doubt by current experiments. However, such an observation is still not conclusive on the origin of the suppression. In particular, data from the Pierre Auger Observatory can be interpreted
in a scenario in which the suppression is due to the maximum acceleration energy at the sources
rather than to interactions in the background radiation. In this scenario, UHECR data can no longer
yield bounds on Lorentz invariance violations which increase the thresholds for interactions of nuclei
on background photons, in particular through modification of the dispersion relations. Here we argue
in turn that the study of UHECRs still represents an opportunity to test Lorentz invariance,
by discussing the possibility of deriving limits on violation parameters from UHECR phenomena
other than propagation. In particular we study the modifications of the shower development in the
atmosphere due to the possible inhibition of the decay of unstable particles, especially neutral pions.
}

\FullConference{The 34th International Cosmic Ray Conference,\\
		30 July- 6 August, 2015\\
		The Hague, The Netherlands}

\begin{document}

\section{Introduction}\label{sec.introduction}
Our universe appears to be very well described by quantum mechanics at
small scales and by general relativity at large scales, but the correct
way to completely unify these two theories is still unknown. It is
possible that the correct theory of Quantum Gravity (QG) would predict
that space-time is subject to quantum fluctuations and the geometry of
the world, with all its symmetries, emerges in the semiclassical limit.
Lorentz Invariance (LI) is therefore not guaranteed to be an exact symmetry, and violations (LIV) can therefore be possible. Although these violations may only be very small, since no measurement has as yet found any evidence, it has been shown in the last two decades that measurable deviations can be present even at energies much lower than the QG scale. In particular, possible LIV effects could show themselves in Ultra High Energy Cosmic Ray (UHECR) phenomena. \\
Although the idea of possible experimental effects of LIV is quite old, the possibility of putting extremely strong limits on, at least some, LIV parameters from UHECRs detection was firstly quantitatively discussed in \cite{Aloisio:2000cm} and later on refined in many ways. It has to be recalled that UHECRs are the most energetic particles in the Universe, and the use of the experimental observations to constrain possible LIV effects is natural, in particular since it has been shown \cite{Greisen:1966jv,Zatsepin:1966jv}, (the Greisen, Zatsepin and Kuzmin - GZK - effect) that these particles do not propagate freely in the Universe but interact with the photon backgrounds (Cosmic Microwave Background Radiation, CMB, and ExtraGalactic Background Light, EGBL) that fill it. The processes involved, photopion production and photo-dissociation, implicate low energies at terrestrial accelerators. On the contrary, they appear to occur at UHE given the small energy of the background photons.\\
Consequently, as soon as the evidence of the suppression in the spectrum of UHECRs around $5 \times 10^{19}$ eV became undisputable, based on results from HiRes \cite{Abbasi:2007sv}, Auger \cite{Abraham:2010mj} and, more recently, Telescope Array \cite{AbuZayyad:2012qk}, limits on those violating parameters were derived. A discussion and references can be found in \cite{Liberati:2013xla}.
All these bounds however rely on the assumption that the observed spectrum is generated in sources which inject nuclei at energies larger than those characteristic of these interactions.\\
Here we discuss the status of these bounds in the light of the combined analysis \cite{ArmandoICRC} of the measurements of the Pierre Auger Observatory both of the spectrum \cite{Abraham:2010mj,ThePierreAuger:2013eja} and nuclear composition \cite{Aab:2014kda} which strongly suggests that the observed suppression in the spectrum might be due to the maximum cosmic ray acceleration energy at the sources rather than to an effect of their propagation in extragalactic space. This fact was recently noted by several authors \cite{Aloisio:2013hya,Taylor:2011ta,Taylor:2015rla,Unger:2015laa,Globus:2015xga}.\\

\section{LIV and UHECR propagation}\label{sec.propagation}
In recent times possible LI violating models have been deeply analyzed and compared with available experimental data \cite{Liberati:2013xla,Kostelecky:2008ts}. In particular, in the approach of Effective Field Theories, possible Lorentz (and CPT) violating terms have been described in the Standard Model Extension (SME) in \cite{Colladay:1998fq}. When symmetry (and renormalizability) is no longer a guide, the number of possible terms is essentially infinite. Those that can be subject to experimental verification (several hundreds) are described in \cite{Kostelecky:2008ts}.\\
To parametrize departures from relativistic invariance, as for instance in \cite{Aloisio:2000cm}, we only consider modification of the dispersion relations (which in the SME approach corresponds to modification of kinetic terms in the Lagrangian),
which amounts to assuming that the relation, connecting the energy and momentum of a particle, is modified as:
\begin{equation}
E_i^2-p_i^2=m_i^2 \Rightarrow \mu^2_i(E,p,M_P)  \approx m_i^2+{\frac{f_i}{ M_P^n}} E_i^{2+n}
\label{eq3}
\end{equation}
where $p=\vert \overrightarrow{p} \vert $, $\mu$ is a function of momenta and energy, $ M_P \approx 1.2 \times 10^{28}$ eV, the Planck mass, is the possible scale where QG effects become important and $f_i$, which can be either positive or negative,  parametrizes the strength of LIV for particle $i$. The last equality reflects the fact that LI is an exceedingly good approximation of the physics we know, so that modifications are expected to be quite small, making an expansion of the LIV dispersion relation in terms of $E/M_P$ appropriate.\\
Here we make the simplifying assumption that all possible terms at the same order have the same form, which amounts to neglect higher order corrections to Eq.~\ref{eq3}. Also, we consider only the lowest correction term in $E/M_P$ present (higher orders are totally negligible at UH energies).
In practical terms, only $n=1,2$ will be relevant \cite{Aloisio:2000cm}.\\ 
The right hand side of Eq.~\ref{eq3} is invariant when $f_i=0$. 
We will assume normal conservation of energy and momentum. In a LIV theory it is always possible to move all the violations to the dispersion relations \cite{Carmona:2012un}.
Finally we assume that, in nuclei, LIV only affects nucleons: since then the relevant energy entering the modified dispersion relations is the energy/nucleon, this implies that violations are milder for nuclei than for protons; for a nucleus of atomic number $A$, effectively  $M_P \rightarrow A M_P$.\\
From  Eq.~\ref{eq3} it is clear that the correction term is always much smaller than both ($E^2,p^2$) even for  $E \approx 10^{20}$ eV. However, as soon as\footnote{Since at the leading LIV order $E \approx p$ we will use them without distinction.} $p \geq (m_i^2 M_P^n /|f_i|)^{1/(2+n)}$ the correction becomes larger than the mass of the particle, and this can lead to very important effects \cite{Aloisio:2000cm}.
We consider here how LIV affects the threshold energy for the GZK process $p \gamma_{bkg} \rightarrow (p,n) \pi$, where $\gamma_{bkg}$ is a photon of the CMB or EGBL. The threshold for this process, in a LIV world, is modified:
\begin{equation}
E_{GZK}\approx \frac{m_p m_{\pi}}{2 \omega_{\gamma}} \Rightarrow
E_{GZK}\approx \frac{\mu(E_p,p_p,m_p,M_P) \mu(E_{\pi},p_{\pi},m_{\pi},M_P)}{2 \omega_{\gamma}}
\label{eq6}
\end{equation} 
($\omega_{\gamma}$ being the energy of the background photon). The last equation has to be solved for $E_p=E_{GZK}$.
For our simplified treatement, we will assume that $f_i$ are the same for all the hadrons.\\
The case we are interested in is for $f_i \le 0$. As soon as $f_i$ moves from zero towards negative values the threshold energy at first slightly increases, but for $f_i <- 2.5 \times 10^{-14}~(n=1)$ [$f_i<- 4 \times 10^{-7}~(n=2)$],  Eq.~\ref{eq6} has no longer real solutions \cite{Aloisio:2000cm}: the photo-pion production reaction is no longer kinematically allowed and protons propagate freely in the Universe, if only interactions on the CMB are taken into account. Interaction on the EGBL are affected for larger values of $-f_i$.\footnote{For positive $f_i$ the thresholds move lo $lower$ values. However other processes are allowed in this case, such as Vacuum Cherenkov $ p \rightarrow p \gamma$, for which very strong bounds exist \cite{Klinkhamer:2008ss}, from the mere existence of UHECRs.}\\
For nuclei,  for which the relevant process of interaction on the universal backgrounds is photo-disintegration, an equation corresponding to Eq.~\ref{eq6}, with  $M_P \rightarrow A M_P$, can be written. The modification of the thresholds is similar to that for protons.\\
Limits on LIV parameters derived from the observed steepening of the spectrum of UHECRs have been reported in literature \cite{Jacobson:2005bg,Saveliev:2011vw}.
These limits,  however, depend crucially on the assumption that the observed flux suppression is originated by the propagation of UHECRs. 
Auger composition data combined with those on the all-particle spectrum 
have been analyzed in \cite{ArmandoICRC} (see also \cite{Aloisio:2013hya,Taylor:2011ta,Taylor:2015rla,Unger:2015laa,Globus:2015xga}) and they indicate, at least when interpreted in a simple source model, a different scenario in which the flux suppression reflects the end of cosmic ray acceleration at the source, with the implication of requiring hard injection spectra.\\
It is therefore worthwhile to verify if LIV can still be bound in this scenario.\\
\begin{figure}[!t]
\hspace{-0.5cm}
\includegraphics*[height=0.5\textwidth]{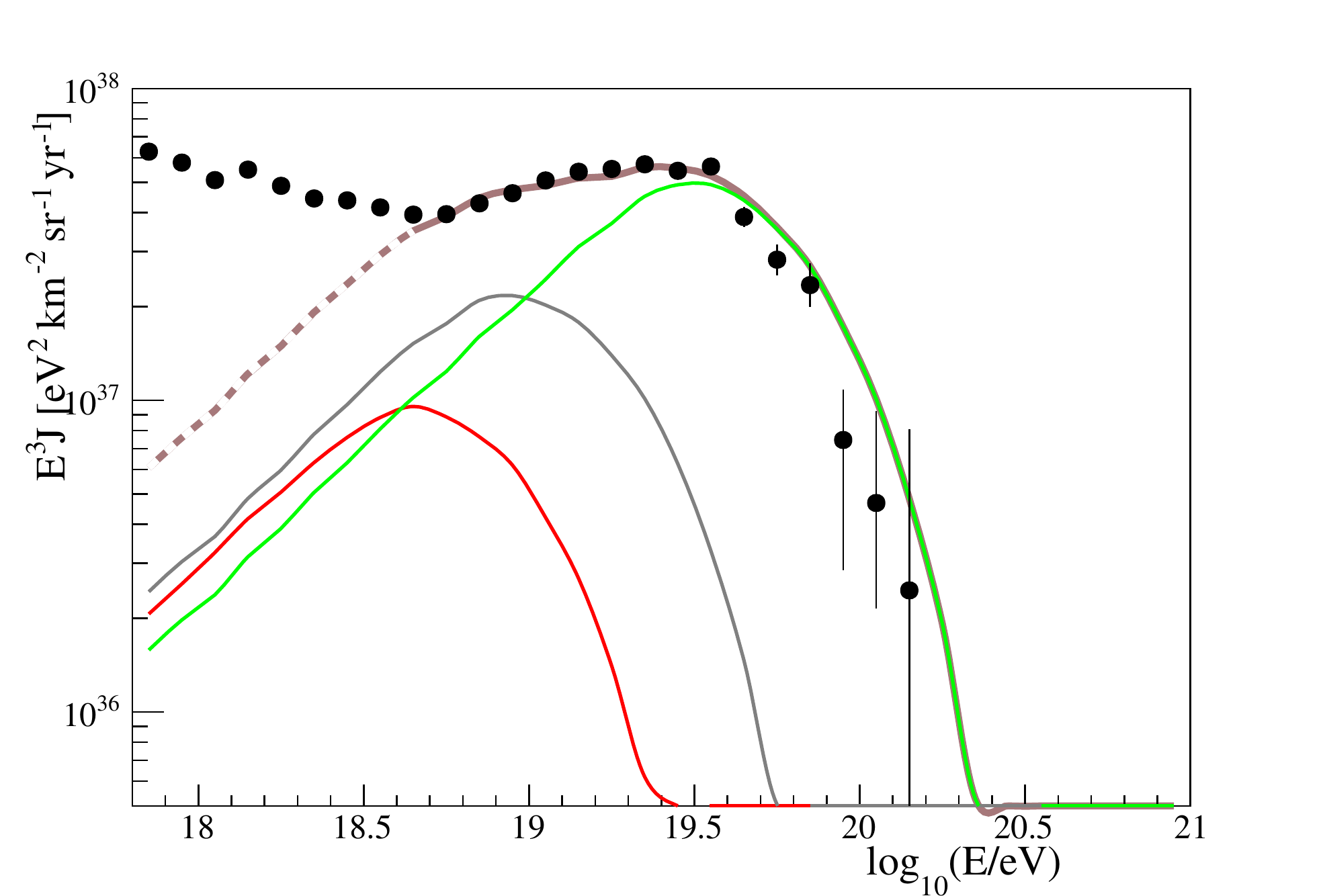}
\hspace{-1.3cm}
\includegraphics*[height=0.5\textwidth]{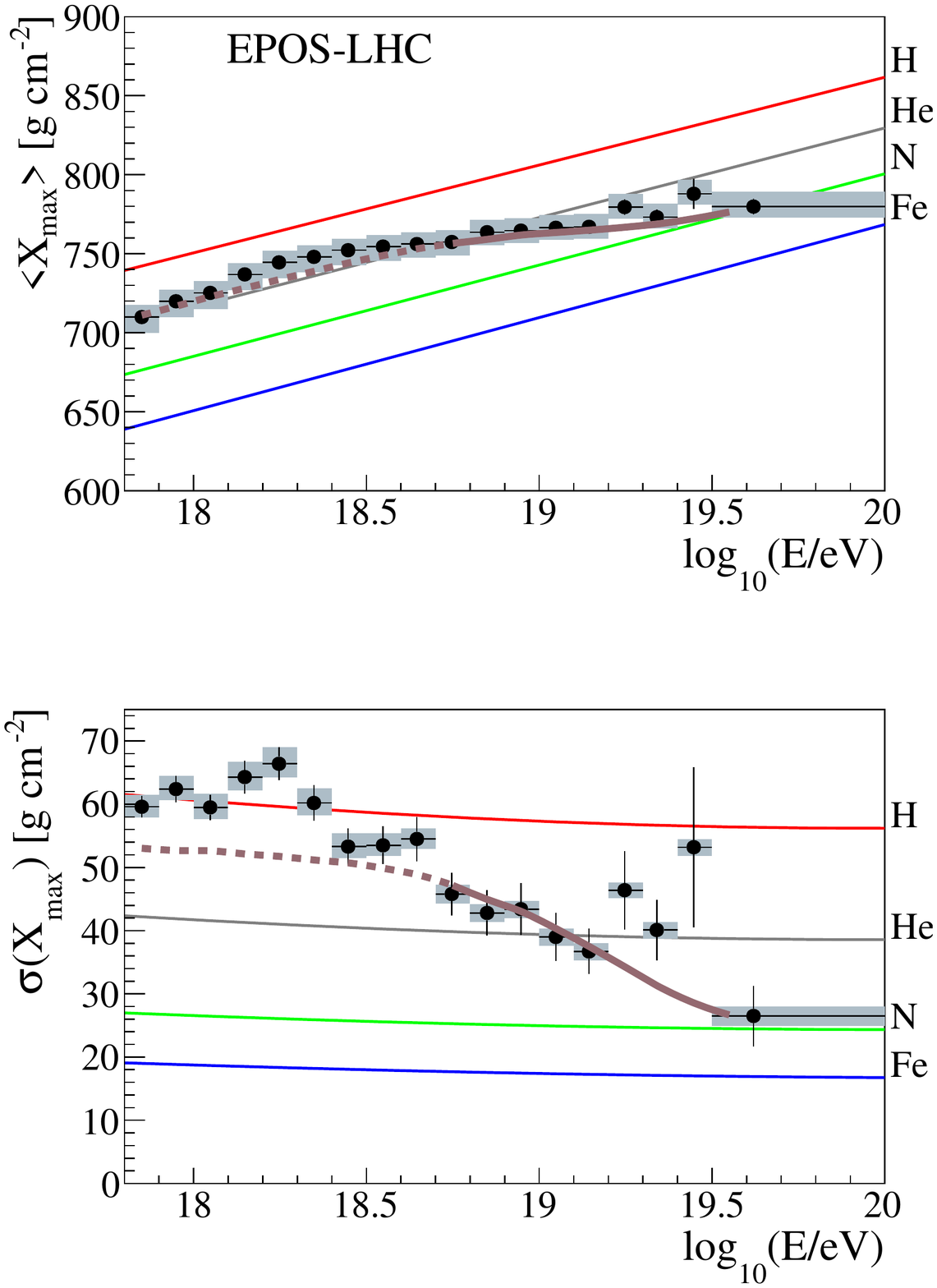} 
\caption{Simulated energy spectrum of UHECRs (multiplied by
$E^3$) at the top of the Earth's atmosphere with maximum source rigidity $R_{max}=5 \times 10^{18} ~V$ and $\gamma=2$, along with Auger data points. The propagation is simulated switching off the interactions with photon backgrounds. Partial spectra are grouped
according to the mass number as follows: $A=1$ (red), $2<A<4$ (grey), $5<A<14$ (green), total (brown). Right: average (top) and standard deviation (bottom) of the $X_{\mathrm{max}}$ distribution (assuming EPOS-LHC for UHECR-air interactions). Gray bands show the systematic uncertainties.}
\label{EposLHCscenario}
\end{figure}
To simulate LIVs we have  propagated  UHECRs switching off the interactions with background photons, only accounting for energy losses due to the expansion of the Universe. To account for these losses we used both a simplified version of {\it SimProp} \cite{Aloisio:2012wj} and a semi-analytical code,
within a simple source model consisting of identical sources, uniformily distributed in comoving volume throughout the whole Universe, emitting all nuclei in a rigidity dependent way. These nuclei are the same impinging on the atmosphere with energy reduced by the redshift. For this analysis we assume injection of 4 nuclei (H, He, N, Fe).
The absence of interactions might make source evolution effects important. We have checked that this does not happen for soft injection spectra as the one discussed here. \\
It has been shown \cite{Aloisio:2012wj,Taylor:2015rla} that it is not possible to reproduce the data with a single extragalactic component, a different one being needed below $E=5 \times 10^{18}$ eV, unless possible primary interactions are considered in the source environment \cite{Unger:2015laa,Globus:2015xga}. For our analysis we will only consider larger energies.\\
We arbitrarily impose fixed $\gamma=2$, consistent with Fermi
acceleration mechanism, and maximum rigidity $R_{cut}=5 \times 10^{18}$ V,   and search the best composition that reproduces the experimental data, taking into account only statistical errors. The results for spectrum and composition are
presented in Fig. \ref{EposLHCscenario}: the resulting composition is light to intermediate (H, He, N). The overall (spectrum and composition) reduced $\chi^2 \approx 2.3$ is acceptable given the simplifications made and does not allow us to exclude LIV  in the propagation of nuclei.\\
It is however obvious that the above statement $cannot$ be taken as evidence of LIV, since many other astrophysical/particle physics explanations can be considered.\\

\section{LIV effects on shower development in the atmosphere}\label{sec.atmosphere}
\begin{figure}[!t]
\centering
\begin{minipage}{.5\textwidth}
  \centering
\includegraphics*[width=1\textwidth]{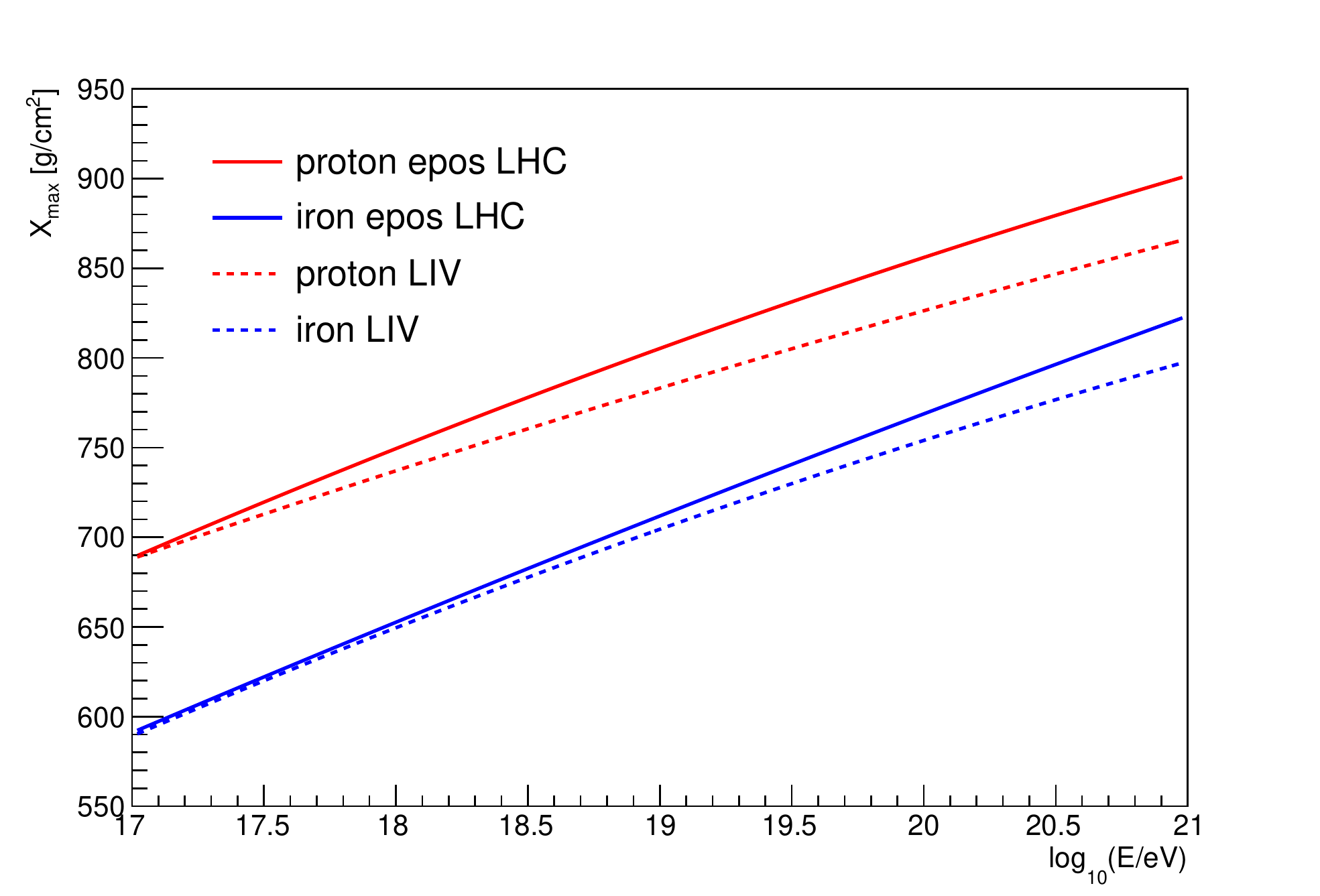}\\
\end{minipage}%
\begin{minipage}{.5\textwidth}
  \centering
  \includegraphics[width=1\linewidth]{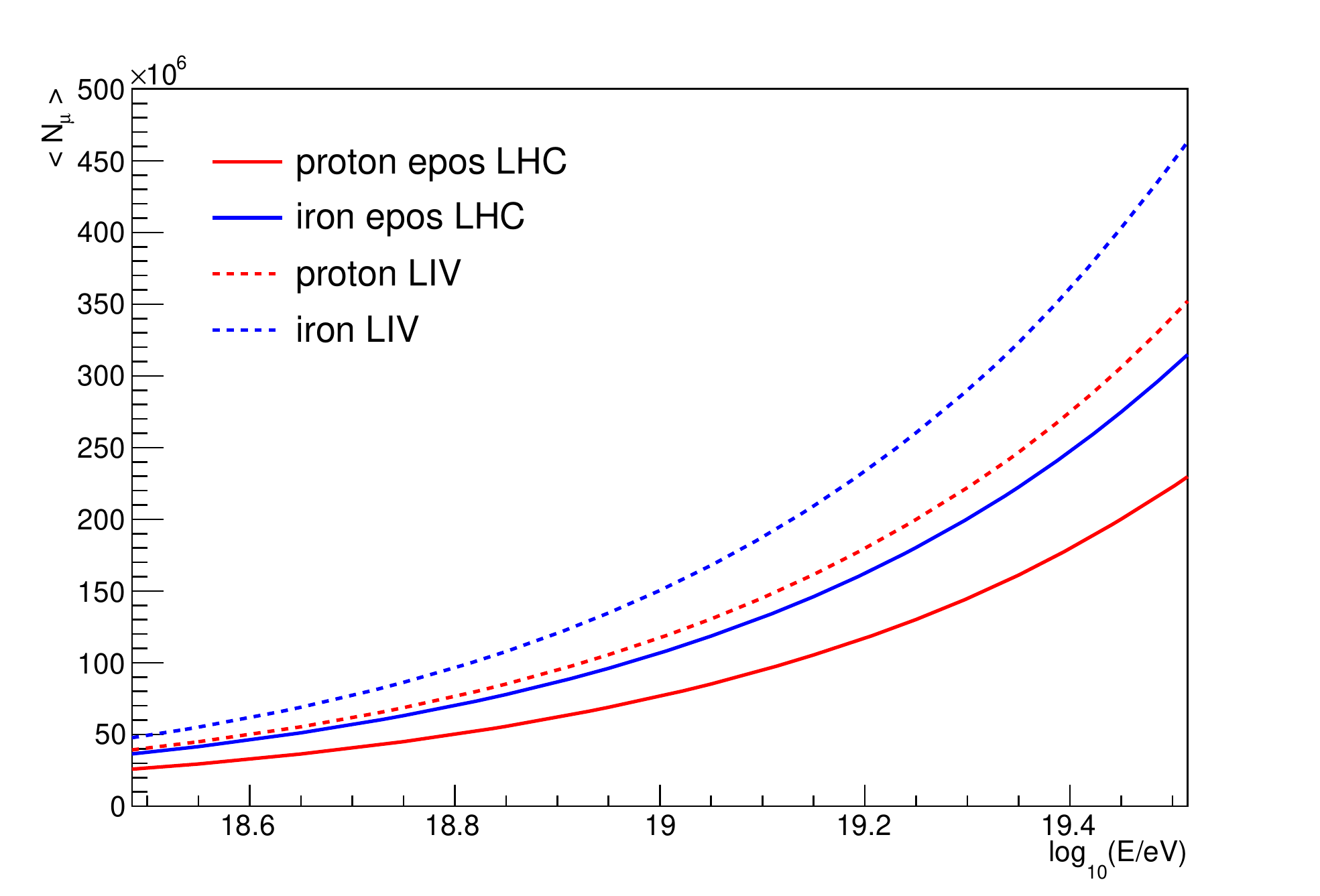}
\end{minipage}
\caption {Left panel: Expectation for $\langle X_{max} \rangle $ vs energy for LI shower development (solid lines) and LIV case (dashed lines). Right panel: Average number of muons vs primary energy in LI and LIV cases.
}
\label{LIVxmaxNmu}
\end{figure}

In principle, $all$ aspects of UHECR physics can be modified by LIV. For instance, LIV can affect the cosmic ray acceleration processes, and  also the energy losses during acceleration. As an example, consider LI violation with $f_i>0$. In this case, as soon as the energy of the accelerated nucleus is above some threshold, vacuum Cherenkov process becomes possible and essentially no further acceleration is possible. This case has been discussed, in a different context, in \cite{Klinkhamer:2008ss}.\\ 
On the other hand important effects are expected in the interactions of UHE particles in the atmosphere and in the decay of secondary particles. These effects can make some parts of the kinematical space forbidden for the processes and therefore make some reaction impossible. \\
Consider the most important decay for atmospheric showering, $\pi^0 \rightarrow \gamma \gamma $. It can be shown that, considering LIV modified dispersion relations, the kinematics of the decay changes into (\cite{Aloisio:2014dua}): 
\begin{equation}
m^2_{\pi}+{\frac{1}{M_P^n}}(f_{\pi}E^{2+n}_{\pi}-f_{\gamma}(E_{\gamma_1}^{2+n}+E_{\gamma_2}^{2+n})) - 2 (E_{\gamma_1}E_{\gamma_2}-p_{\gamma_1} p_{\gamma_2}) =2 p_{\gamma_1} p_{\gamma_2} (1-\cos \theta_{1,2})
\label{eq12}
\end{equation}
Since there are very strong limits \cite{Vankov:2002gt} on $f_{\gamma}$ we will assume it to be zero.
The right hand side of Eq.~\ref{eq12} is non negative, while the left hand one can become negative for large enough $E_{\pi}$ and $f_{\pi}<0$. Therefore neutral pions do not decay if
$ E_{\pi}>(M_P^n m_{\pi}^2/|f_{\pi}|)^{\frac{1}{2+n}}\approx 6/|f_{\pi}|^{1/3} \times 10^{15} $ eV ($n=1$) ($\approx 1.4/|f_{\pi}|^{1/4} \times 10^{18}$ eV, $n=2$). The case with $n=2$ is expected to give a very small effect and so it will not be considered in the following.\\
To test this effect we have generated $10^5$ atmospheric showers with a modified version of CONEX \cite{Bergmann:2006yz} in which the
decay of unstable particles is inhibited for $ E_i \geq (M_P m_i^2/|f_i |)^{1/3}$ fixing $-f_i=1$. \footnote{For the case considered, the effect of LIV can be opposite for particles with respect to antiparticles. This effect has been neglected in the present work.} The results of this simulation are presented in Figs.~\ref{LIVxmaxNmu}. In the left panel, the
expectation for $X_{\mathrm{max}}$ vs energy for LI shower development (solid lines) and LIV
case (dashed lines) is reported. As one can see, the change in the particles'decay
(mainly neutral pions) has the net effect to move the shower maximum to higher
altitudes as the electromagnetic part of the shower consumes faster. From the
observational point of view the LIV effects on the shower development make the
interactions of nuclei (and protons) primaries compatible with (LI) interactions of
heavier nuclei.\\
\begin{figure}[!t]
\hspace{-0.5cm}
\includegraphics*[height=0.5\textwidth]{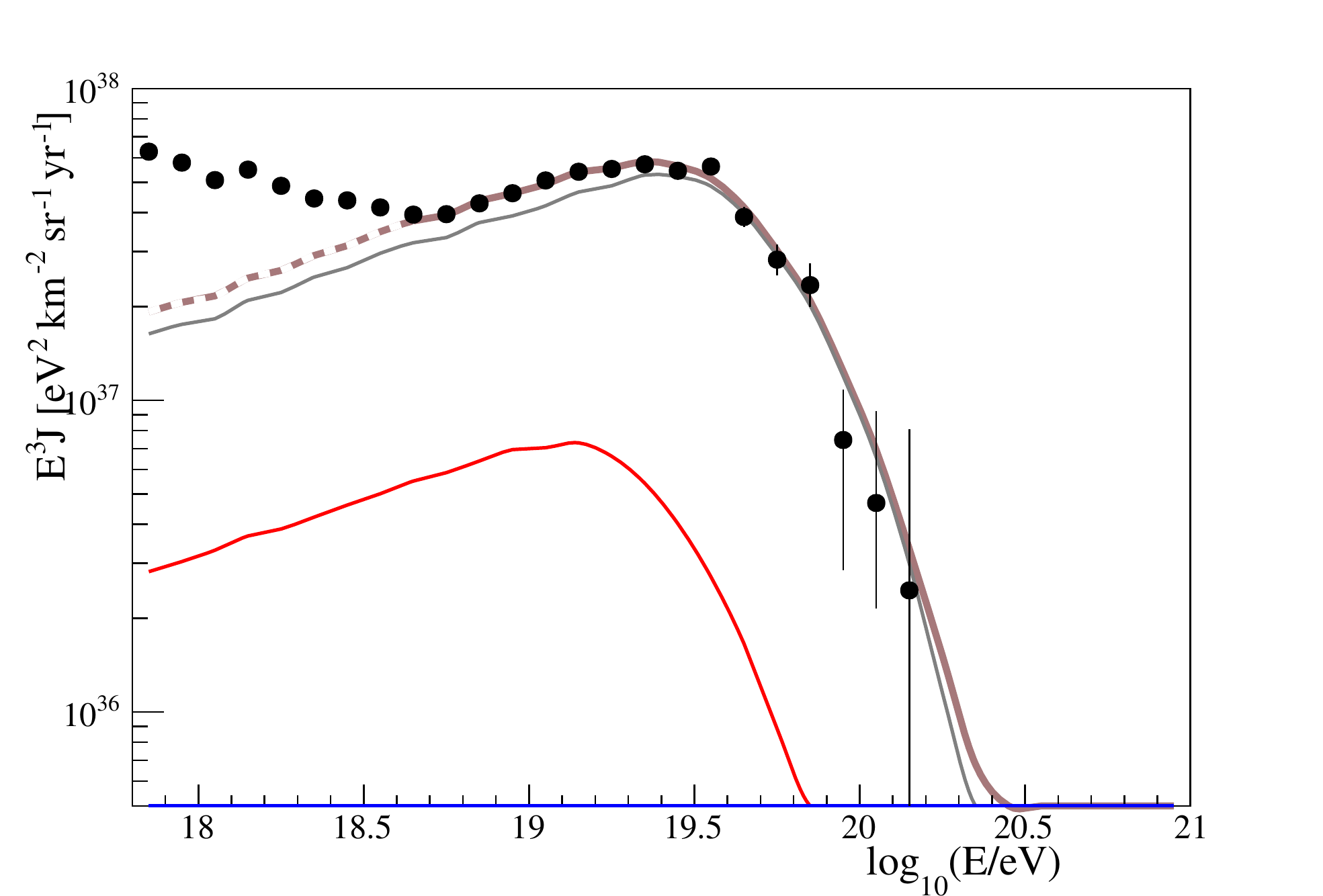}
\hspace{-1.3cm}
\includegraphics*[height=0.5\textwidth]{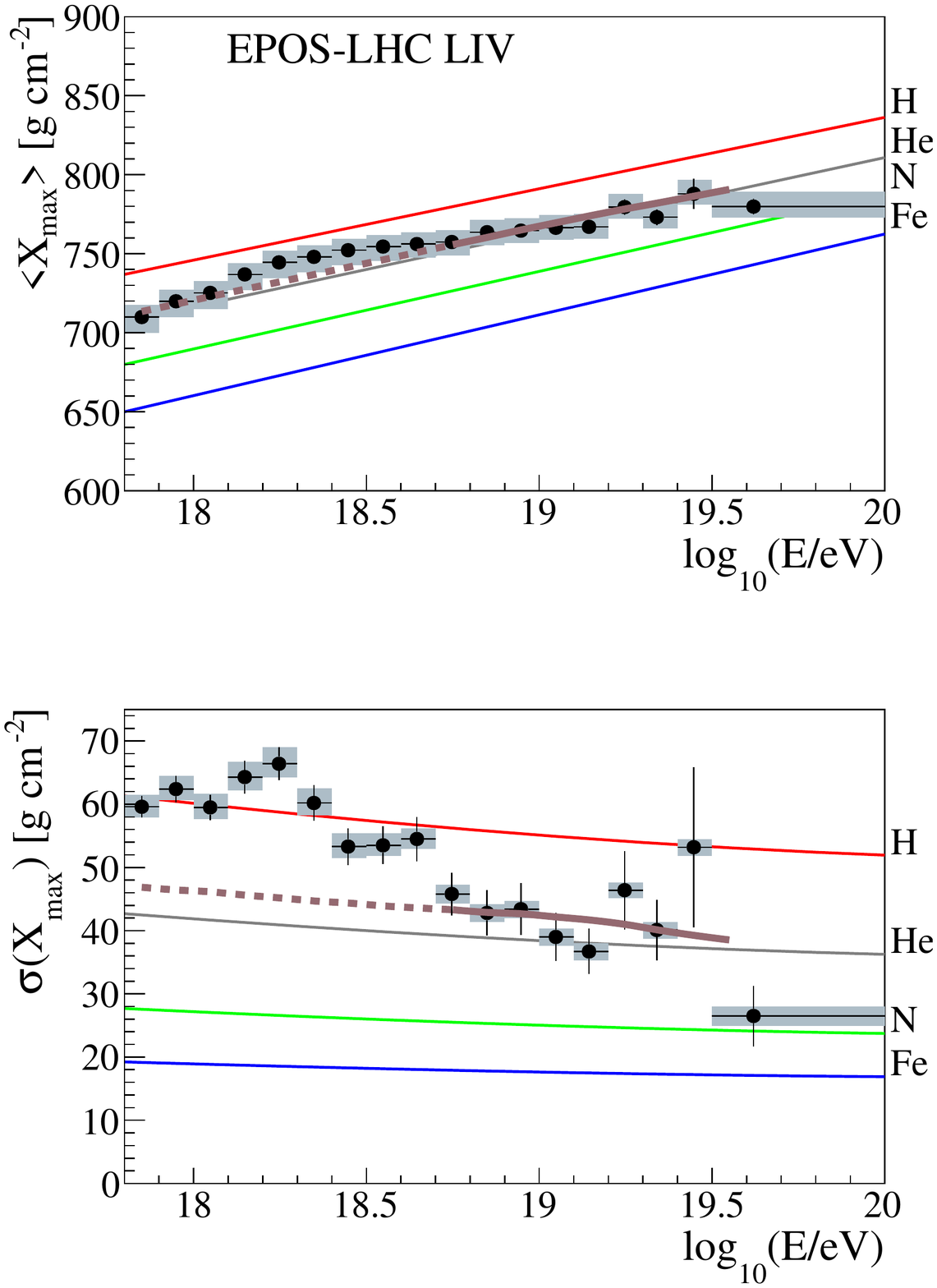} 
\caption{Same as in Fig.~1. In this cas a maximum source rigidity $R_{max}= 10^{19.2} ~V$ and $\gamma=2.65$ are used. Here EPOS for UHECR-air interactions has been modified so as to take into account the LIV effects.}
\label{EposLIVscenario}
\end{figure}
We incidentally note that another effect of the suppression of neutral pion decays
at high energies would be the increase of the number of their interactions. This
would give more muons at ground level with respect to LI shower development, as
shown in Fig.~\ref{LIVxmaxNmu} (right panel) where the average number of muons vs primary
energy in LI and LIV cases is reported. A study of possible constraints to LIV from observational data on
muons in showers will be addressed in a forthcoming analysis.\\
The modifications of the shower development due to LIV (Fig.~\ref{LIVxmaxNmu}, left panel) do affect the
analysis of the spectrum and composition data. Including such modification, the solution reported
in Fig.~\ref{EposLHCscenario} corresponds to a reduced $\chi^2 > 4$. However a better
description of the data is found at $\gamma = 2.65$, $R_{cut} \approx 1.6 \times 10^{19}$ V with a reduced $\chi^2 \approx 1.6$. In this case, reported in
Fig. \ref{EposLIVscenario}, the composition changes and is dominated by light elements (H, He): there is in fact no need of heavier primaries, given the change
of the interactions in the atmosphere.
It has to be reminded, however, that the modifications of the shower developement described here are only relevant for the case of rather strong violations {\it i.e.} $n=1$ and $f_i \approx O(1)$. Milder violations give smaller effects. Similar effects in a LI context were presented in \cite{Allen:2013hfa}.

\section{Conclusions}\label{sec.conclusions}
UHECRs are the highest energy particles since almost immediately after
the Big Bang. It is therefore natural to use their propagation and
interactions to probe the structure of space-time, and in particular
possible tiny violations of Lorentz Invariance. Experimental data, both on the spectrum and chemical composition, from
the Pierre Auger Observatory have been recently analyzed and the
suppression of the all particle spectrum has been interpreted as a
possible manifestation of the maximum energy reached at sources that
accelerate nuclei in a rigidity dependent way. \\
We have analyzed the status of limits on LIV in this scenario,
assuming violation in the propagation through the universal background
radiation: for values of the violation parameter $f_i <- 2.5 \times 10^{-14}~(f_i<- 4 \times 10^{-7})$, interactions on photons are switched off, still
experimental data can be reasonably described, so that essentially no
bounds on LIV parameters can be cast.\\
Violations can also affect the interaction of UHECRs in the atmosphere
and in particular the shower development. We have found that even in
this case we cannot exclude this effect, that is however much weaker
than that relevant for propagation. 
The results reported here cannot be taken as evidence for LIV. Even for the hard injection spectra astrophysical explanations are available (see for instance \cite{Kotera:2015pya} and references therein), and (non LIV) changes in the interactions at the highest energies are possible. 
However, finer spectrum and composition measurements at the highest energies can conceivably be used to put
constraints on (at least some) violations of relativistic invariance.\\
Finally we remark that we have discussed here only a small number of cases of all the possible violations. These are probably the most important affecting UHECR propagation and interaction, but certainly a general analysis of {\it{all}} aspects of LIV on cosmic ray physics would be very welcome.

\section*{Acknoledgements}
We thank our Auger colleagues for discussions. The research of DB is supported by Progetto Speciale Multiasse ``La Societ{\`a} della Conoscenza in Abruzzo'' PO FSE Abruzzo 2007 - 2013.

\bibliography{LIVbib}
\bibliographystyle{JHEP}

\end{document}